\documentclass[aps,prx,twocolumn,superscriptaddress]{revtex4-2}

\usepackage{amsmath}
\usepackage{amsfonts}
\usepackage{amssymb}
\usepackage{dcolumn}
\usepackage[english]{babel}
\usepackage{epsfig}
\usepackage{natbib}
\usepackage{graphics}
\usepackage{xcolor}

\usepackage{placeins}
\usepackage{comment}
\usepackage{dsfont}
\usepackage{braket}
\usepackage{amsmath}

\makeatletter
\renewcommand{\fnum@figure}{Fig. \thefigure}
\makeatother

\newcommand{\dd}{\mathrm{d}}
\newcommand{\ee}{\mathrm{e}}
\newcommand{\ii}{\mathrm{i}}
\newcommand{\abs}[1]{\left| #1 \right|}

\DeclareMathOperator*{\sumsum}{\sum\sum}
\newcommand{\myparallel}{{\mkern3mu\vphantom{\perp}\vrule depth 0pt\mkern2mu\vrule depth 0pt\mkern3mu}}

\begin{document}

\title{Angular Bloch Oscillations and their applications}

\author{Bernd Konrad}
\affiliation{German Aerospace Center (DLR), Institute of Quantum Technologies, 89081 Ulm, Germany}

\author{Maxim Efremov}
\affiliation{German Aerospace Center (DLR), Institute of Quantum Technologies, 89081 Ulm, Germany}

\date{\today}

\begin{abstract}

To advance precise inertial navigation, we present a compact quantum sensor which is based on novel quantum phenomenon of the angular Bloch oscillations and measures solely the angular acceleration of slow external rotation.
We investigate the dynamics of ultra-cold atoms confined in a toroidal trap with a ring-lattice along the azimuth angle, realized with the superposition of two copropagating Laguerre-Gaussian beams.
In the presence of external rotation of small angular acceleration, or prescribed linear chirp between the two beams, the measured angular momentum of trapped atoms displays a specific periodic behaviour in time, which we name as the angular Bloch oscillations.
This discovered quantum phenomenon is shown to be a key element of  fruitful applications for (i) an efficient transfer of quantized angular momentum from light field to atoms by controlling the chirp, and (ii) the direct determination of the angular acceleration of external rotation by measuring the Bloch period.

\end{abstract}

\maketitle

\section{Introduction} 

Bloch oscillations (BOs) describe the periodic motion of a particle in a periodic potential responding to a constant force. 
Initially proposed for electrons in the periodic crystal structure of metals by Bloch in 1929~\cite{Bloch1929}, this quantum phenomenon remained non-observable for many decades, because the Bloch period, being inversely proportional to the lattice period, exceeds the decoherence time in the system by many orders. 
However, in atomic physics the lattice period could be substantially extended and BOs have been observed with utracold atomic gas trapped in optical lattice~\cite{Dahan1996, Peik1997}. 
The fundamental fact that the atom picks up the linear momentum from the light during BOs has improved substantially the sensitivity and precision of atom interferometers.
Nowadays they are utilized in many metrological applications, including precise measurements of fine structure constant~\cite{Clade2011}, gravitational acceleration~\cite{Charriere2012, Geiger2020, Abend2016}, gravity gradient~\cite{Kasevich2002,Asenbaum2017}, the Newtonian gravitational constant~\cite{Lamporesi2008,Rosi2014}, and the Coriolis non-inertial force~\cite{Pritchard1997,Kasevich1997,Gauguet2009,Stockton2011,Berg2015,Dutta2016,Durfee2006,Wu2007} 
The latter is of great importance for realizing large enclosed space-time areas~\cite{Muller2012} and observing interference of massive objects~\cite{Fein2020}.  

In addition, the field of inertial navigation~\cite{Nusbaum2019} and seismic records~\cite{MatIsa1992} require sensors for precise measurement of the time-dependent angular velocity. At present, there have already been developed indirect~\cite{Schreiber2023, Ovaska1998} and direct~\cite{Nusbaum2019} schemes to determine the angular acceleration. 
In particular, when the angular velocity changes faster than the repetition rate of measurement, the angular acceleration can be estimated with an array of classical inertial sensors~\cite{Nilsson2016}. 

In this paper, we make a prediction of new quantum phenomenon of the angular Bloch oscillations (ABOs) and present valuable applications of ABOs to (i) transfer efficiently the {\it quantized} portion of angular momenta to the particle and (ii) realize {\it compact quantum} sensor for measuring exclusively the angular acceleration of external rotation. 
To observe ABOs in a lab, we propose and consider in detailed the experimental scheme, which is based on ultra-cold atomic gas tightly confined in a toroidal trap with an additional periodic potential along the angular variable. 
Such a ring lattice can be realized via superposition of two copropagating Laguerre-Gaussian (LG) beams with the angular indices $l$ and $-l$~\cite{FrankeArnold2007}. We show that in the presence of external rotation of constant angular acceleration, or prescribed linear chirp between the LG beams, the measured angular momentum of trapped atoms displays the characteristic periodic behavior in time, that is ABOs. 
As the first application, we proof that ABOs are mostly prominent in the regime of a shallow ring lattice and thus offer a very efficient way to transfer the well-defined amount of angular momenta, namely multiple of $2l\hbar$, from the field to the trapped atoms. 
This technique significantly advances guided-wave Sagnac atom interferometers~\cite{Navez2016} and persistent current in a toroidal Bose-Einstein condensate~\cite{Phillips2007, Ramanathan2011}. 
As the second application, we demonstrate that our setup based on ABOs measures in fact only the non-inertial Euler force, being entirely determined by the angular acceleration, rather than the Coriolis or centrifugal ones. 

Our paper is organized as follows.
In Section~\ref{sec:setup} we introduce our setup for realizing a ring lattice suitable for observation of ABOs. Here we also show how the strong confinement of ultra-cold atoms allows us to describe dynamics appearing in the proposed scheme by the effective one-dimensional model.   
We proof in Section~\ref{sec:ABOS} that the resulting one-dimensional system indeed displays ABOs and then present the detailed timeline of observation procedure.
Section~\ref{sec:Applications} presents fruitful applications of ABOs, namely to transfer the quantized angular momentum from the ring lattice to the atoms, calibrate experimentally our effective one-dimensional model and design quantum sensor for determining the angular acceleration of external rotation. 
We briefly summarize in Section~\ref{sec:Conclusion} our results and provide an outlook. 
In order to keep our article self-contained but focused on the central ideas, many lengthy calculations are included in four appendices. 
In Appendix~\ref{sec:AppendixAdiabatic} we investigate the adiabatic time evolution of our system. 
The loading of atoms into the ground-state of a ring lattice is quantitatively studied in Appendix~\ref{sec:AppendixLoadingProcess}. 
We then consider in Appendix~\ref{sec:AppendixLZ} the non-adiabatic transitions induced by fast rotation. 
Finally, Appendix~\ref{sec:AppendixApproximateGroundStateBand} is devoted to derivation of the analytical formulas for the first two energy bands in the case of a shallow lattice. These results help us to evaluate analytically the main parameters of ABOs.

\section{ Setup}

\label{sec:setup}

\subsection{Three-dimensional scheme}

\begin{figure}[t]
	\centering
	\includegraphics[width=\columnwidth]{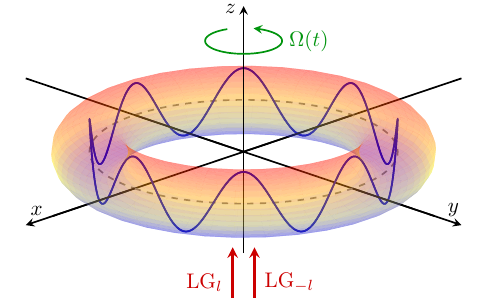}
	\caption{ 
		Setup for observing angular Bloch oscillations.
		Ultracold atoms are held by the toroidal trap, restricting their movement on the ring, dashed line, with radius $r_0$ at $z=0$.
		Two red-detuned Laguerre-Gaussian beams with opposing azimuthal indices $\pm l$ are used to create the ring lattice potential, blue line, which can be moved along the ring via the frequency chirp.
		Moreover, the atoms experience the external rotation $\Omega (t)$ around the symmetry axis of the ring.
	} 
	\label{fig:Setup} 
\end{figure}

For observing angular Bloch oscillations (ABOs), we consider the following setup. 
A gas of ultracold atoms of mass $m$ is confined in the three-dimensional (3D) trap described by the total potential~\cite{Amico2021,AtomtronicRMP2022}
\begin{align}
	\label{eq:total potential}
	U(\mathbf{r},t) =U_\mathrm{T}(\mathbf{r})  + U_\mathrm{RL}(\mathbf{r},t),
\end{align}
where the first term $U_\mathrm{T}({\mathbf{r}})$ corresponds to the toroidal trap
\begin{align}
	U_\mathrm{T}(\mathbf{r}) = \frac{m \omega_\perp^2}{2} \left( r-r_0 \right)^2
	+
	\frac{m \omega_\myparallel^2}{2} z^2,
	\label{eq:SetupPotentialTorus}
\end{align}
keeping atoms on a specific ring with radius $r_0$ and in the plane $z=0$, as shown in Fig.~\ref{fig:Setup}. 
Here we use the cylindrical coordinates $(r, \varphi, z)$ for the position vector ${\mathbf r}$ and denote the trapping frequency along the $r$ and $z$ direction by $\omega_\perp$ and $\omega_\myparallel$, accordingly. 

The second term $U_\mathrm{RL}(\mathbf{r},t)$ in Eq.~\eqref{eq:total potential} describes the ring-lattice trap
\begin{align}
	U_\mathrm{RL}(\mathbf{r},t) =U f_l\left[\frac{\sqrt{2}r}{w(z)} \right]\left\{ \cos^2\left[ l\varphi - a(t) \right] -1 \right\},
	\label{eq:SetupPotentialLGBeams}
\end{align}
providing a periodic modulation along the angular variable $\varphi$ with the amplitude $U f_l [\sqrt{2}r/w(z)]$ depending on $r$ and $z$. 
Here, $l=1,2,\ldots$ and $w(z)$ is the beam waist, $w(z) = w_0 \sqrt{1+(z/z_\mathrm{R})^2}$ with the Rayleigh length $z_\mathrm{R}$.

The potential $U_\mathrm{RL}(\mathbf{r})$ is an optical dipole potential realized with the linear superposition of two red-detuned LG optical beams $\mathrm{LG}_l$ and $\mathrm{LG}_{-l}$, copropagating along the $z$-axis, as depicted in Fig.~\ref{fig:Setup}. 
These LG beams have the same intensity $I_\mathrm{LG}$ and the radial index $p=0$, but opposite azimuthal indices $l$ and $-l$. 
Thus, we have $f_l(\xi) = \exp(-\xi^2) \xi^{2l}/l!$ and $U=(1/\pi)(\hbar \Gamma^2/|\Delta|)(I_\mathrm{LG}/I_\mathrm{sat})$, where $\Gamma$ denotes the natural line width of the used atomic transition, $\Delta<0$ the laser detuning, and $I_\mathrm{sat}$ the saturation intensity. 
Moreover, the position-dependent relative phase $({\Delta\omega(t)}/{c})\left[z-{r^2}/{R(z)}\right]$, with $\Delta\omega(t) = \omega^{(-l)}(t) - \omega^{(l)}(t)$ and $R(z)$ being the radius of curvature of the beam wavefront, is typically small~\cite{FrankeArnold2007} and therefore can be neglected. 

In addition, to rotate the ring lattice, these LG beams are phase-locked and have the relative chirp $\Delta\omega(t)$. Indeed, this chirp defines the function
\begin{align}
	a(t) = \frac{1}{2}\int_{t_0}^t \!\Delta \omega(t')\,\dd t',
	\label{eq:SetupDefinitionChirp}
\end{align}
determining the positions of maxima and minima of the ring-lattice potential $U_\mathrm{RL}(\mathbf{r},t)$, according to Eq.~\eqref{eq:SetupPotentialLGBeams}.

\subsection{From 3D to effective 1D}
\label{sec:SetupDimensions}

When we neglect atom-atom interaction, the dynamics of atoms in the potential $U(\mathbf{r},t)$ is governed by the 3D Schr\"odinger equation
\begin{align}
	\ii \hbar \frac{\partial}{\partial t} \Psi
	=
	\left[
	-\frac{\hbar^2}{2m} \frac{\partial^2}{\partial \mathbf{r}^2}
	+U(\mathbf{r},t)
	\right] \Psi
	\label{eq:Setup3DDynamics}
\end{align}
for the wave function $\Psi(\mathbf{r},t)$. 

As is well known, the amplitude of the linear BOs reaches its maximum in the regime of the shallow lattice. 
Consequently, to study the Bloch oscillations, now on the finite domain of the ring (effectively 1D system), our experimental setup has to be designed in a way that the atoms are {\it strongly} confined on the ring, $r=r_0$ and $z=0$, by the toroidal trap, Eq.~\eqref{eq:SetupPotentialTorus}, and experience only {\it weak} periodic potential along the angle $\varphi$ provided by the ring-lattice trap, Eq.~\eqref{eq:SetupPotentialLGBeams}. 
To achieve this, we first require that the characteristic length of confinement $\sqrt{\hbar/(m\omega_\myparallel)}$ along $z$ is much smaller than the Rayleigh length $z_\mathrm{R}$, $\sqrt{\hbar/(m\omega_\myparallel)}\ll z_\mathrm{R}$. 
This effectively allows us to consider the ring-lattice potential $U_\mathrm{RL}$, Eq.~\eqref{eq:SetupPotentialLGBeams}, entirely at $z=0$. 
Moreover, since the function $f_l(\xi)$ has a single maximum at $\xi=\sqrt{l}$, $f_l(\sqrt{2}r/w_0)$ in $U_\mathrm{RL}(\mathbf{r},t)$ has a maximum at $r=\sqrt{l/2} w_0$. 
This radius needs to be matched with the minimum of the toroidal trap $U_\mathrm{T}$, Eq.~\eqref{eq:SetupPotentialTorus}, at $r=r_0$, resulting in the condition $r_0=\sqrt{l/2} w_0$. 
We also demand that the toroidal trap creates a well-defined ring structure, that is the ring width $\sqrt{\hbar/(m\omega_\perp)}$ is small compared to its radius $r_0$, $\sqrt{\hbar/(m\omega_\perp)}\ll r_0$.

As a result, when the frequencies $\omega_\perp$ and $\omega_\myparallel$ in the radial and axial directions are sufficiently large in comparison with the characteristic frequency $\omega_\varphi\sim \hbar/(mr_0^2)$ in the angular direction, the entire dynamics takes place only in $\varphi$ direction. 
In this way, the wave function $\Psi(\mathbf{r},t)$ can then be approximated by
\begin{align}
	\label{eq:wave-function}
	\Psi(r,\varphi,z,t) = Z_0(z)\frac{R_0(r)}{\sqrt{r}}\Phi(\varphi,t)\ee^{-\frac{\ii}{\hbar} \varepsilon_0 t},
\end{align}
where $Z_0(z)$ and $R_0(r)$ represent the normalized wave-functions of the ground state only in the toroidal trap $U_\mathrm{T}(\mathbf{r})$, Eq.~\eqref{eq:SetupPotentialTorus}, in the $r$ and $z$ directions, accordingly, $\varepsilon_0$ is the energy phase factor. 
The wave-function $\Phi(\varphi,t)$ describes the dynamics in the $\varphi$ direction and obeys the effective 1D Schr\"odinger equation
\begin{align}
	\ii \hbar \frac{\partial}{\partial t} \Phi = 
	\left\{ 
	-\frac{\hbar^2}{2I} \frac{\partial^2}{\partial \varphi^2}
	+ V \cos^2 \left[ l\varphi -a(t) \right] 
	\right\} \Phi
	\label{eq:Setup1DReducedDynamics}
\end{align} 
with the effective moment of inertia
\begin{align}
	I =& \left[ 
	\int_0^\infty \! \dd r \frac{\abs{R_0(r)}^2}{mr^2}\,
	\right]^{-1}
	\label{eq:ReductionFormulaforI}
\end{align}
and the effective potential depth
\begin{align}
	V = U \!
	\int_0^\infty \! \dd r
	\int_{-\infty}^\infty \! \dd z
	\abs{Z_0(z)}^2
	\abs{R_0(r)}^2 f_l\left[ \frac{\sqrt{2}r}{w(z)} \right].
	\label{eq:ReductionFormulaforV}
\end{align}

The energy phase factor $\varepsilon_0$ in Eq.~\eqref{eq:wave-function} then reads 
\begin{align}
	\varepsilon_0
	=
	\int_0^\infty \! \dd r\frac{\abs{R_0(r)}^2}{4r^2} \,
	+\frac{\hbar}{2}\left( \omega_\myparallel + \omega_\perp \right)
	-V.
\end{align}

For sufficiently strong toroidal trap, the parameters $I$, Eq.~\eqref{eq:ReductionFormulaforI}, and $V$, Eq.~\eqref{eq:ReductionFormulaforV}, of our 1D effective model, described by Eq.~\eqref{eq:Setup1DReducedDynamics}, can be easily estimated, $I\approx {mr_0^2}$ and $V\approx U f_l(\sqrt{2}r_0/w_0)= U f_l(\sqrt{l})= U (l^l/l!)\ee^{-l}$. 
However, the exact values of $I$ and $V$ are unknown and depend on specific details of the actual experimental setup. In Section~\ref{sec:Parameters} we show how the ABOs can be applied for the determination or calibration of these parameters experimentally. 

In fact, there are also other possibilities to create a similar system. 
Any periodic modulation along the ring (the $\varphi$ direction) leads to the same results we are presenting in this paper. In particular, one can use the blue-detuned LG beams instead, which builds up a dark lattice, where the atoms are trapped in the vicinity of the small intensity. This gives rise to longer coherence times due to smaller scattering rates~\cite{FrankeArnold2007}. In this case, the correct combination of the azimuth indices $l_1$ and $l_2$ of the LG beams should be found to create a ring lattice~\cite{Malinovsky2023}. The number of sites is still equal to $\abs{l_2-l_1}$.

\section{ Angular Bloch Oscillations}

\label{sec:ABOS}

The previous section has presented the experimental scheme to realize a ring lattice suitable for observation of ABOs. 
In this section, we first show that the resulting 1D system indeed displays ABOs and then discuss in detail an observation procedure for this phenomenon.

\subsection{Particle in a ring lattice}

\label{sec:TimeEvo1D}

If there is additionally an external rotation with the time-dependent angular velocity $\mathbf{\Omega}(t) = \Omega (t) \mathbf{e}_z$ around the symmetry axis of the ring, Fig.~\ref{fig:Setup}, the effective 1D dynamics of our system is described by the Schr\"odinger equation 
\begin{align}
	\ii \hbar \frac{\partial}{\partial t} \Phi = 
	\left\{ 
	-\frac{\hbar^2}{2I} \frac{\partial^2}{\partial\varphi^2}
	+
	V\cos^2 \left[ l\varphi -a(t) \right] 
	+
	\Omega(t) \ii \hbar \frac{\partial}{\partial\varphi} 
	\right\} \Phi
	\label{eq:1DDynamicsMasterEquation}
\end{align} 
for the wave function $\Phi (\varphi,t)$. This equation should be solved on the finite domain $-\pi < \varphi \leq \pi$ with periodic boundary condition for $\Phi (\varphi,t)$, namely $\Phi(-\pi,t)=\Phi(\pi,t)$ and $[\dd\Phi/\dd\varphi](-\pi,t)=[\dd\Phi/\dd\varphi](\pi,t)$.

In general, the dynamics of our system, Eq.~\eqref{eq:1DDynamicsMasterEquation}, is controlled by three parameters: 
$(\mathrm{i})$ the magnitude of the periodic potential $V(t)$,
$(\mathrm{ii})$ the function $a(t)$, and $(\mathrm{iii})$ the angular velocity $\Omega(t)$ of external rotation.
Now we show that the effect of the chirp can be treated as effective rotation. 
When looking at any point of the periodic potential, \textit{e.g.} a maximum or a minimum, its angular position moves with the angular velocity $\dot{a}(t)/l$, where we denote the time derivative of $a(t)$ as $\dot{a}(t) = \dd a/\dd t$. 
Therefore, by considering our system in the co-rotation frame, which moves with the angular velocity $\dot{a}(t)/l$ with respect to the lab frame, the transformed wave function 
\begin{align}
	\label{eq:Theta-Phi}
	\Theta(\varphi,t) = \exp{\left[\frac{a (t)}{l} \frac{\partial}{\partial\varphi}\right]} \Phi(\varphi,t)
\end{align}
obeys the new form of the Schr\"odinger equation
\begin{equation}
	\ii \hbar \frac{\partial}{\partial t} \Theta = \left[ -\frac{\hbar^2}{2 I} \frac{\partial^2}{\partial\varphi^2} + V(t)\cos^2(l\varphi) +\Omega_{\rm eff}(t)\ii \hbar \frac{\partial}{\partial\varphi} \right]\Theta,
	\label{eq:1DdynamicsRotating}
\end{equation}
which should be solved on the domain $\varphi \in (-\pi,\pi]$  with the periodic boundary conditions. 
Here the total time-dependent angular velocity
\begin{equation}
	\Omega_{\rm eff}(t) \equiv \Omega(t)+\frac{\dot{a}(t)}{l}=\Omega(t)+\frac{\Delta\omega(t)}{2l}
	\label{eq:Omega_eff_definition}
\end{equation}
combines together two sources of rotation and defines
\begin{align}
	-\hbar\eta (t) \equiv I\Omega_{\rm eff}(t)
	\label{eq:DefEta}
\end{align}
as the total angular momentum associated with this rotation of the atoms on the ring. 
As a result, the dynamics in the co-rotating frame is entirely determined only by the two parameters: $V(t)$ and $\Omega_{\rm eff}(t)$.

\subsection{Bloch states in a non-accelerating ring lattice}

Before studying the dynamics of atoms in a ring lattice, we first investigate this system in the case of the constant potential amplitude $V(t)=V$ and rotation with the constant angular momentum $\hbar \eta (t)= \hbar\eta_0$. 
The right-hand side of Eq.~\eqref{eq:1DdynamicsRotating} is time-independent and periodic in the angular variable $\varphi$ with period $\Delta \varphi = \pi/l$. 
According to the Floquet theorem, the solutions of Eq.~\eqref{eq:1DdynamicsRotating} then read
\begin{equation}
	\label{eq:1DdynamicsRotating-Solution}
	\Theta(\varphi, t)=\Theta_{n,m}(\varphi)\exp\left(-\ii E_{n,m}^{\Theta}t/\hbar\right),
\end{equation}
where $E_{n,m}^{\Theta}$ is the eigenenergy and the corresponding wave-functions of the Bloch states
\begin{align}
	\Theta_{n,m}(\varphi)= \ee^{\ii m\varphi} u_{n,m+\eta_0}(\varphi)
	\label{eq:EigenstatesBloch}
\end{align}
always consist of a plane wave $\exp \left({\ii m\varphi}\right)$ and the periodic function $u_{n,m+\eta_0} (\varphi)$ with the same periodicity $\Delta \varphi$ as the potential. 
Here $m$ is restricted to integers from the first Brillouin zone $m \in (-l,l]$, defining the corresponding recoil energy and frequency
\begin{align}
	E_\mathrm{r} \equiv \hbar \omega_\mathrm{r} \equiv \frac{\hbar^2l^2}{2I},
\end{align}
while $n$ numbers the band index, $n=0, 1, 2,\ldots$~\cite{Peden2007,Peden2010}.
Consequently, our system displays the band structure with $2l$ states per band.

The wave-functions $u_{n,m+\eta_0}(\varphi)$ are solutions of the ordinary differential equation
\begin{align}
	{E}_n(q) u_{n,q} = 
	\left[
	\frac{1}{2 I} \left(-\ii \hbar\frac{\dd}{\dd\varphi} + \hbar q\right)^2+V \cos^2(l\varphi)
	\right]
	u_{n,q},
	\label{eq:Ustationarysolutions}
\end{align}
which should be solved for $-\pi/2l < \varphi \leq \pi/2l$ with periodic boundary conditions, that is $u_{n,q}(-\pi/2l)=u_{n,q}(\pi/2l)$ and $[\dd u_{n,q}/\dd\varphi](-\pi/2l)=[\dd u_{n,q}/\dd\varphi](\pi/2l)$, and $q=m+\eta_0$. 

Moreover, the energies $E_{n,m}^{\Theta}$ in Eq.~\eqref{eq:1DdynamicsRotating-Solution} can be obtain from ${E}_n(q)$ by using the relation
\begin{align}
	\label{eq:Energy_relation}
	{E}^{\Theta}_{n,m} = {E}_n(m+\eta_0)-\frac{\hbar^2 \eta_0^2}{2I}.
\end{align}

\begin{figure}
	\centering
	\includegraphics[width=\columnwidth]{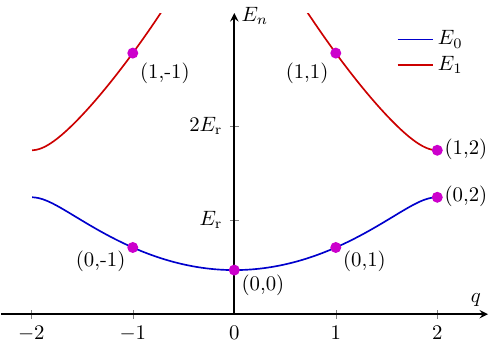}
	\caption{ 
		The energy bands $E_0(q)$, blue curve, and $E_1(q)$, red curve, for $V=E_\mathrm{r}$ and $l=2$. 
		The purple dots labeled by the indices $(n,m)$ display the discrete energies $E_n(m)$ of the system for vanishing angular momentum, $\hbar \eta_0=0$.
	} 
	\label{fig:EnergyDisp} 
\end{figure}

In Fig.~\ref{fig:EnergyDisp}, we display the energies $E_0(q)$ (blue curve) and $E_1(q)$ (red curve) for the first Brillouin zone, $q\in(-l,l]$, which have been obtained by solving Eq.~\eqref{eq:Ustationarysolutions} numerically for $V=E_\mathrm{r}$ and $l=2$. 
The periodicity of the function $u_{n,q}(\varphi)$ in $\varphi$ automatically implies that the eigenenergies $E_n(q)$ are periodic in $q$, that is ${E}_n(q+2l) = {E}_n(q)$. 
In addition, the purple dots labeled with double index $(n,m)$ represent the discrete energies ${E}_n(m)$ of our system in the case of vanishing angular momentum, $\hbar \eta_0=0$, as given by Eq.~\eqref{eq:Energy_relation}. 

\begin{figure}
	\centering
	\includegraphics[width=\columnwidth]{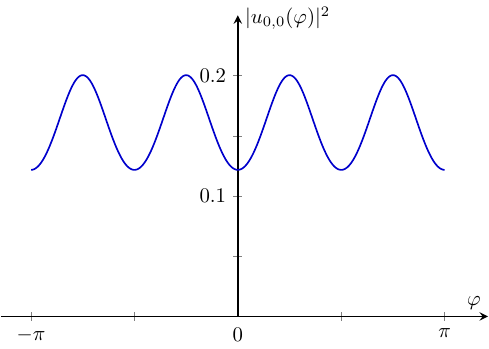}
	\caption{  
		Probability distribution $\abs{u_{0,0}(\varphi)}^2$ of the ground state of the system for $V=E_\mathrm{r}$, $l=2$ and $ \hbar \eta_0=0$.
	} 
	\label{fig:GSEnergyDisp} 
\end{figure}

Finally, in Fig.~\ref{fig:GSEnergyDisp}, we present the normalized distribution $|u_{0,0}(\varphi)|^2$ of the ground state, corresponding to $m=0$ and $n=0$.

\subsection{ABOs in an accelerating ring lattice}
\label{sec:AcceleratingLattice}

\begin{figure}[t]
	\centering
	\includegraphics[width=\columnwidth]{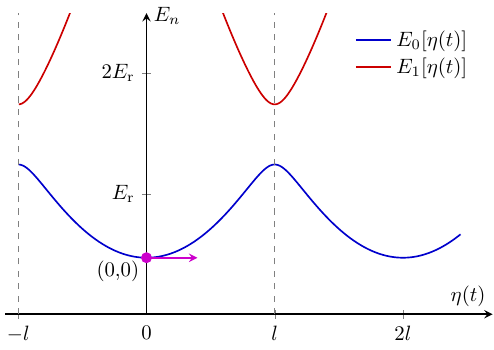}
	\caption{  
		Angular Bloch oscillations viewed with the energy dispersion. 
		The system is prepared in its initial ground state, purple dot, and follows the lowest-energy band $E_0(q)$, blue line, during adiabatic dynamics.
	} 
	\label{fig:ABOsEnergyDisp} 
\end{figure}

Now we investigate the dynamics appearing in a ring lattice rotating with the constant angular acceleration $\dot{\Omega}_\mathrm{eff}=\dd \Omega_{\mathrm{eff}}/\dd t$, corresponding to the linear change of the total angular momentum $\eta(t)=st$ in time, with $s=-I\dot{\Omega}_\mathrm{eff}/\hbar$, Eq.~\eqref{eq:DefEta}.  If the system is initialized in its ground state with the wave-function $u_{0,0}(\varphi)$ and the energy $E_{0,0}^\Theta$, as depicted by the purple dot in Fig.~\ref{fig:ABOsEnergyDisp}, for small enough $s$, the system is staying in its instantaneous eigenstate and moves along the blue solid line, following the energy dispersion relation, as shown in Fig.~\ref{fig:ABOsEnergyDisp}.

According to Appendix~\ref{sec:AppendixAdiabatic}, such an adiabatic dynamics is described mathematically by the approximate solution   
\begin{align}
	\Theta(\varphi,t) = \ee^{\ii \phi(t)} u_{0, \eta(t)}(\varphi)
	\label{eq:AdiabaticTimeEvo}
\end{align}
of Eq.~\eqref{eq:1DdynamicsRotating}, where the imprinted time-dependent phase
\begin{align}
	\phi(t) = \phi_\mathrm{d}(t) + \phi_\mathrm{g}(t)
\end{align}
consists of the dynamical phase
\begin{align}
	\phi_\mathrm{d}(t) = 
	-
	\frac{1}{\hbar}
	\int_{0}^t \! {E}_{0, 0 }^{\Theta} (st')\, \dd t'
	\label{eq:DynamicalPhase}
\end{align}
and the geometrical phase
\begin{align}
	\phi_\mathrm{g}(t) = \ii \int_{0}^t \! 
	\Bra{u_{0,\eta(t')}} \frac{\partial}{\partial t'} \Ket{u_{0, \eta(t')}}
	\, \dd t',
	\label{eq:GeometricalPhase}
\end{align}
originating from the change of the instantaneous eigenstates $\Ket{u_{0, \eta(t')}}$ with time \cite{Schleich}.

Since the energy dispersion $E_0(\eta)$ is a periodic function, $E_0(\eta+2l)=E_0(\eta)$, as displayed in Fig.~\ref{fig:ABOsEnergyDisp}, we intuitively expect the periodic dynamics in our system after the Bloch time 
\begin{align}
	t_\mathrm{B} = \frac{2l}{s}.
	\label{eq:BlochTime}
\end{align}

Indeed, evaluating the wave-function $\Theta(\varphi,t)$, Eq.~\eqref{eq:AdiabaticTimeEvo}, after a time step of $t_\mathrm{B}$
\begin{align}
	\Theta(\varphi,t+t_\mathrm{B}) = \ee^{\ii \delta\phi(t)}
	\ee^{-\ii 2l \varphi}
	\Theta(\varphi,t),
	\label{eq:BlochTimeStep}
\end{align}
we arrive at the original wave-function $\Theta(\varphi,t)$ multiplied by the time-dependent global phase $\delta\phi(t) = \phi(t + t_\mathrm{B})- \phi(t)$, as well as the $\varphi$-dependent phase $-2l\varphi$, describing the increase  of the angular momentum by $2\hbar l$. Here we have used the fact that $u_{0,\eta(t)+2l}=\exp(-2\ii l\varphi)u_{0,\eta(t)}$. 

Thus, our system indeed displays periodic dynamics in the state and we call this phenomenon the angular Bloch oscillations. Now, the important question arises: which measurable quantity does display this periodic behaviour? 
For the conventional Bloch oscillations in a linear periodic potential~\cite{Peik1997}, such a periodic behavior occurs in the mean velocity. 
In analogy, we analyze the mean value $ \braket{\hat{L}_z}\!(t)=\Bra{\Phi(t)} \hat{L}_z \Ket{\Phi(t)}$ of the angular momentum operator $\hat{L}_z$ as the function of time, where $\Phi(\varphi,t)=\exp[- \ii a\hat{L}_z/(\hbar l)]\Theta(\varphi,t)$ is the wave-function in the lab frame. 
With $\Theta(\varphi,t)$ given by Eq.~\eqref{eq:AdiabaticTimeEvo}, we obtain
\begin{align}
	\braket{\hat{L}_z}\!(t) = -\hbar \eta(t)+\frac{I}{\hbar} \left. \frac{\dd {E}_0(q)}{\dd q} \right|_{q =\eta(t)}. 
	\label{eq:lzMeasurement}
\end{align}

\begin{figure}[t]
	\centering
	\includegraphics[width=\columnwidth]{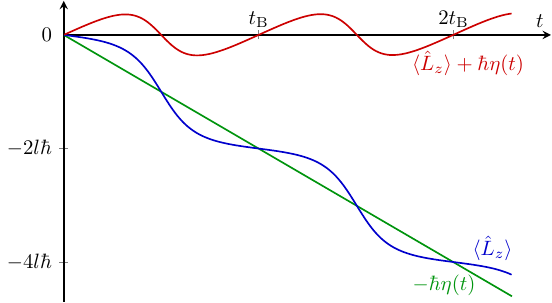}
	\caption{Mean angular momentum of the atoms as the function of rotation time $t$, blue line, for the lattice depth $V=3 E_\mathrm{r}$ and the angular acceleration $s=0.01 \omega_\mathrm{r}$. 
		In addition, the linear behaviour $-\hbar \eta(t)=-\hbar st$ is shown by the green line, while the function $\braket{\hat{L}_z}\!(t) + \hbar\eta( t)$ is displayed by the red line.
	} 
	\label{fig:ABONormal} 
\end{figure}

During the adiabatic evolution, $\braket{\hat{L}_z}\!(t)$ has two contributions: $(\mathrm{i})$ the total angular momentum $\hbar\eta(t)$, picked up by the state, and $(\mathrm{ii})$~the derivative of the energy band ${E}_0(q)$ evaluated at $\hbar\eta(t)$. 
In Fig.~\ref{fig:ABONormal} we present the dependence $\braket{\hat{L}_z}\!(t)$, blue line, together with the linear shift $-\hbar \eta(t)=-\hbar s t$, green line, for the lattice depth $V = 3 E_\mathrm{r}$ and the scanning parameter $s=0.01\,\omega_\mathrm{r}$. 
When we subtract $-\hbar\eta(t)$ from $\braket{\hat{L}_z}\!(t)$, we obtain only the contribution from the second term $(I/\hbar)[\dd E_0(q)/\dd q](\eta(t))$ on the right-hand side of Eq.~\eqref{eq:lzMeasurement}. 
This function $\braket{\hat{L}_z}\!(t) + \hbar\eta( t)$ of $t$ is displayed in Fig.~\ref{fig:ABONormal} by red line and indeed is the periodic one with the Bloch period $t_\mathrm{B}$, Eq.~\eqref{eq:BlochTime}.
This periodicity entirely originates from the periodicity of the band structure $E_n(q)$ ($n=0,1,2\ldots$) with respect to $q$, $E_n(q+2l)=E_n(q)$, and can be employed to measure the parameters $I$ and $V$ of the effective 1D model of the ring lattice, as shown in Section~\ref{sec:Parameters}.

\subsection{Scheme for observation of ABOs}
\label{sec:Procedure}

\begin{figure}
	\centering
	\includegraphics[width=\columnwidth]{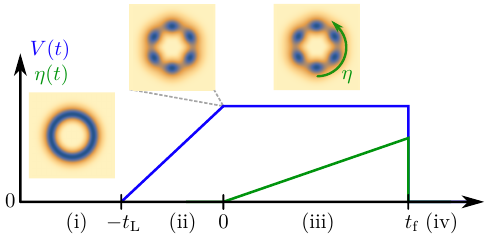}
	\caption{Timeline of the proposed scheme to observe ABOs. The procedure consists of four steps: $\mathrm{(i)}$~state preparation, $\mathrm{(ii)}$~loading, $\mathrm{(iii)}$~rotation and $\mathrm{(iv)}$~measurement. 
		The magnitude $V(t)$ of the ring lattice and the angular velocity $\eta(t)$  are shown as functions of time $t$ by the solid blue and green lines, accordingly. The pictures display snapshots of the trapping potential at the specific times, where blue regions correspond to deeper trapping potential.  
	} 
	\label{fig:Procedure} 
\end{figure}

In order to observe ABOs identified in the mean value $\braket{\hat{L}_z}\!(t)$, we propose a four-step procedure:
$\mathrm{(i)}$~state preparation, $\mathrm{(ii)}$~loading, $\mathrm{(iii)}$~rotation and $\mathrm{(iv)}$~measurement.

\textit{First of all}, the trapped atoms are prepared in the ground state of the toroidal potential $U_\mathrm{T}(\mathbf{r})$, Eq.~\eqref{eq:SetupPotentialTorus}. 
The angular part of this state $\Phi(\varphi,- t_\mathrm{L})=1/\sqrt{2\pi}$ corresponds to the ground state of a particle on a ring.
Fig.~\ref{fig:Procedure} displays the toroidal trap in the $x$-$y$ plane for $t<-t_\mathrm{L}$, marked by the index $\mathrm{(i)}$, where the blue color represents the region of the deeper potential. 

\textit{At the second step}, the ground state of the total potential $U_\mathrm{T}(\mathbf{r})+U_\mathrm{RL}(\mathbf{r})$ should be prepared. 
For this purpose, by switching on slowly the intensity of the LG beams, we are able to increase adiabatically the magnitude $V(t)$ of angular modulation during the loading time $t_\mathrm{L}$, as depicted in Fig.~\ref{fig:Procedure}. In this way, the ground state of the toroidal trap is slowly transferred into the one of the total potential, that is $\Phi(\varphi,0) = u_{0,0}(\varphi)$. Appendix~\ref{sec:AppendixLoadingProcess} provides more details on the required adiabacity of this loading process.
Additionally, we show in Fig.~\ref{fig:Procedure} the final form of the trapping potential at the end of the loading process marked by the index (ii).

\textit{At the third step}, we keep the intensity of the LG beams constant, while having a linear dependency of the relative chirp $\Delta\omega(t)$ between the beams in time. 
This corresponds to the time interval $0<t<t_\mathrm{f}$ when $\eta(t)=s t$, shown by the green solid line in Fig.~\ref{fig:Procedure}, leading to a ring-lattice potential rotating with the constant angular acceleration $\dot{\Omega}_\mathrm{eff}=-\hbar s/I$, Eq.~\eqref{eq:DefEta}.
Such rotation should be again done in an adiabatic manner, that is for small enough $s$, to avoid any transitions from the lowest energy band to the other ones. In Appendix~\ref{sec:AppendixLZ} we investigate precisely the limitations on the ramp parameter $s$. 

Finally, \textit{at the fourth step}, we measure the angular momentum of the state after keeping the constant acceleration for the time $t_\mathrm{f}$. This is carried out by turning off the ring-lattice potential $U_\mathrm{RL}(\mathbf{r})$ abruptly, as shown in Fig.~\ref{fig:Procedure}, and measuring the angular momentum at $t \geq t_\mathrm{f}$. This is possible because the angular momentum is a constant of motion for a particle in the toroidal trap.

Thus, when the adiabatic conditions are fulfilled, the presented procedure can be performed with different rotation times $t_\mathrm{f}$ for measuring $\braket{\hat{L}_z}\!(t_\mathrm{f})$ and therefore observing the angular Bloch oscillations, Fig.~\ref{fig:ABONormal}.

\section{Applications}
\label{sec:Applications}

We have discovered the new phenomenon of the ABOs and presented in detail the scheme to observe them in a lab. Now we focus on three applications of this phenomenon. Here, we first discuss the method to transfer the large {\it quantized} angular momentum from the ring lattice to the atoms. We further utilize ABOs for the direct experimental {\it calibration} of our effective 1D model and then for the {\it sensing} of angular acceleration. 

\subsection{Large angular momentum transfer}
\label{sec:ApplLAMT}

With our setup, ABOs can be used as a tool for transferring a large angular momentum to the center-of-mass degree of freedom of the atom. This process is similar to the one of the conventional Bloch oscillations on the line, where the atom picks up the large and well-defined linear momentum~\cite{Abend-book}.

First of all, ABOs occur only when the ramp parameter $s\equiv (\dd\eta/\dd t)$ of the effective angular acceleration of the ring lattice is quite small, $s \ll s_\mathrm{c}$, where
\begin{align}
	s_\mathrm{c} = \frac{\pi l}{\hbar} \frac{V^2}{ 32 E_\mathrm{r}}.
	\label{eq:ApplicationsScrit}
\end{align}

As shown in Appendix~\ref{sec:AppendixLZ}, under this condition, there occur no the Landau-Zener transitions and the atoms follow adiabatically the slowly rotating ring lattice, with the atomic wave-function in co-rotation frame given by Eq.~\eqref{eq:AdiabaticTimeEvo}. Hence, according to Eq.~\eqref{eq:BlochTimeStep}, for a given rotation time $t$, the average angular momentum $\braket{\hat{L}_z}\!(t)$ changes by      
\begin{equation}
	\braket{\hat{L}_z}\!(t+t_\mathrm{B}) -
	\braket{\hat{L}_z}\!(t)=-2l\hbar
	\label{eq:ABOLzSteps}
\end{equation}
after each Bloch period $t_{\rm B}$. 

\begin{figure}
	\centering
	\includegraphics[width=\columnwidth]{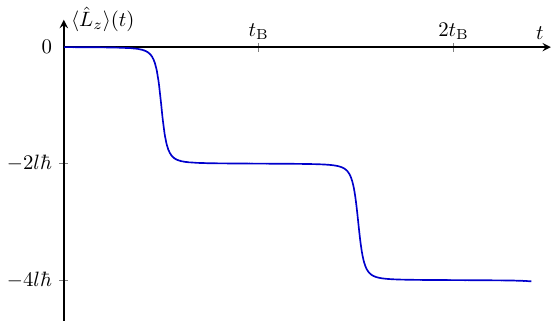}
	\caption{
		Mean angular momentum as the function of rotation time $t$, for lattice magnitude $V=0.5 E_\mathrm{r}$ and the angular acceleration $s=0.002 \omega_\mathrm{r}$. 
	} 
	\label{fig:LargeAngularMomentumTransfer} 
\end{figure}

The quantized transfer of $\braket{ \hat{L}_z} \! (t)$ becomes even more prominent in the case of shallow lattices, as clearly seen from the comparison of Fig.~\ref{fig:ABONormal}, for $V=3 E_\mathrm{r}$, and Fig.~\ref{fig:LargeAngularMomentumTransfer}, for $V=0.5 E_\mathrm{r}$. 
For weaker lattices, the step-like change of $\braket{ \hat{L}_z} \! (t)$ takes place only at the edges of the Brillouin zones, $t=(n+1/2)t_{\mathrm{B}}$ with $n=0,1,2\ldots$, while $\braket{ \hat{L}_z} \! (t)$ is almost constant at other times. However, the weaker lattice reduces $s_{\mathrm{c}}$, Eq.~\eqref{eq:ApplicationsScrit}, and thereby increases the experimental time needed for keeping adiabatic time evolution.

Still, we have to keep in mind that these results are obtained within the effective 1D model of our proposed setup, Eq.~\eqref{eq:1DDynamicsMasterEquation}, where there occurs only the dynamics in the $\varphi$ direction. However, for quite large transferred angular momentum, the centrifugal energy becomes comparable to the excitation energy $\hbar\omega_\perp$, leading to the energy transition between the $\phi$ to $r$ directions. This process limits the maximally allowed transfer of the angular momentum from the ring lattice to the atoms.

\subsection{Parameters of the effective 1D model}
\label{sec:Parameters}

Now we take a look back at the effective 1D dynamics governed by Eq.~\eqref{eq:1DDynamicsMasterEquation}. 
Despite the external rotation and the chirp, we have two parameters characterizing the system, namely the effective moment of inertia $I$ and the magnitude $V$ of the effective 1D periodic potential. Depending on the specific setup used, the reduction from the 3D description to the effective 1D model provides approximate values of $I$, Eq.~\eqref{eq:ReductionFormulaforI}, and $V$, Eq.~\eqref{eq:ReductionFormulaforV}. For any application of ABOs, it is therefore extremely useful to measure these values directly and even compare them to the analytical results. 

In order to calibrate $I$ and $V$, we consider no external rotation, $\Omega(t)=0$, and prescribe the chirp $\Delta \omega (t)$ in a linear manner,
\begin{align}
	\Delta \omega (t) = -{\mathcal B} t
	\label{eq:ChirpWithB}
\end{align}
with a well-controlled constant ${\mathcal B}$. 

\begin{figure}
	\centering
	\includegraphics[width=\columnwidth]{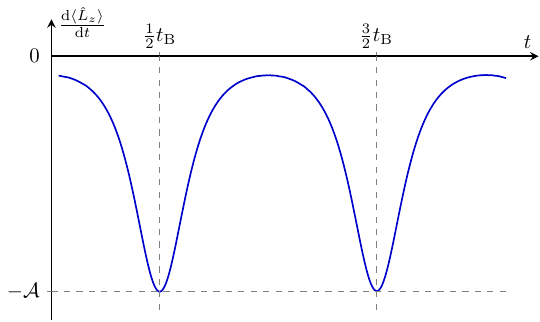}
	\caption{Time derivative of the mean angular momentum $\langle \hat{L}_z\rangle (t)$,  for a lattice depth $V=0.5 E_\mathrm{r}$ and the angular acceleration $s=0.002 \omega_\mathrm{r}$.
	} 
	\label{fig:ABODeriv} 
\end{figure}

In the setup proposed in Section~\ref{sec:setup}, this corresponds to the total angular velocity $\Omega_\mathrm{eff}(t) = -{\mathcal B}t/(2l)$ in Eq.~\eqref{eq:1DdynamicsRotating} and the constant angular acceleration $\dot{\Omega}_\mathrm{eff}=-{\mathcal B}/(2l)$. As shown in Section~\ref{sec:ABOS}, this chirp induces ABOs in our system. Thus, to extract both parameters $I$ and $V$, we take the derivative of $\braket{ \hat{L}_z}\! (t)$ with respect to $t$ and obtain $\dd\!\braket{ \hat{L}_z}\!/\dd t$, presented in Fig.~\ref{fig:ABODeriv}. 
This curve displays the periodic behavior with the Bloch period $t_\mathrm{B}$ and has sharp peaks at $t=(n+1/2)t_{\mathrm{B}}$ with $n=0,1,2\ldots$, corresponding to the edges of the Brillouin zones. The distance between two neighbouring peaks measures the Bloch period $t_\mathrm{B}$. By using Eq.~\eqref{eq:BlochTime} with $s=(\dd\eta/\dd t)$ and Eq.~\eqref{eq:DefEta}, we derive the relation between the Bloch period
\begin{equation}
	\label{eq:tB_application}
	t_\mathrm{B}= 4l^2\frac{\hbar}{I{\mathcal B}}
\end{equation}
and the moment of inertia $I$ for given $l$ and ${\mathcal B}$. 

Next, according to Appendix~\ref{sec:AppendixApproximateGroundStateBand}, the magnitude $\mathcal{A}$ of these peaks, as indicated in Fig.~\ref{fig:ABODeriv}, depends not only on $I$ and ${\mathcal B}$, but also on the potential magnitude $V$, namely in the regime of shallow lattice, $V\ll E_\mathrm{r}$, we have derived
\begin{equation}
	\mathcal{A}= 2l\frac{\hbar^2 {\mathcal B}}{V}
	.
	\label{eq:PeakHeightA}
\end{equation}
Additionally, these peaks are shown to be sharp, since the relative width
\begin{align}
	\frac{\Delta t_\mathrm{FWHM}}{t_\mathrm{B}} = \frac{V}{8 E_\mathrm{r}}\sqrt{4^{1/3}-1}
\end{align}
is small for shallow lattices, where $\Delta t_\mathrm{FWHM}$ denotes the full width at half maximum of the peaks. 

As a result, these two Equations~\eqref{eq:tB_application} and~\eqref{eq:PeakHeightA} allow us to calibrate the parameters $I$ and $V$, 
\begin{align}
	I&= 4l^2\frac{\hbar}{{\mathcal B} t_\mathrm{B}}
\end{align}
and
\begin{align}
	V&= 2l\frac{\hbar^2 {\mathcal B}}{\mathcal{A}}
\end{align}
via measuring the two parameters $t_\mathrm{B}$ and ${\mathcal A}$ of ABOs. The latter are only induced by the prescribed chirp $\Delta\omega(t)$, Eq.~\eqref{eq:ChirpWithB}, in the case of no external rotation.

\subsection{Measurement of external angular acceleration}

Due to the equivalence of external rotation and the chirp between the LG beams in our setup, we can employ ABOs for sensing the external rotation. 
Indeed, according to Eqs.~\eqref{eq:1DdynamicsRotating} and~\eqref{eq:Omega_eff_definition}, in the co-rotating frame the atoms are trapped in the non-rotating periodic potential $V\cos^2(l\varphi)$ and experience total rotation with the effective angular velocity $\Omega_\mathrm{eff}(t)$. 

\begin{figure}
	\centering
	\includegraphics[width=\columnwidth]{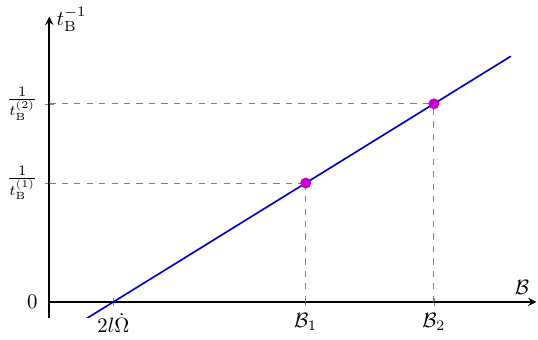}
	\caption{
		Dependence of the inverse Bloch time $t_\mathrm{B}^{-1}$, Eq.~\eqref{eq:InvBlochTimeChirpRotation}, on the prescribed chirp constant ${\mathcal B}$.
	} 
	\label{fig:Application2ExtAA} 
\end{figure}

For external rotation with the constant acceleration $\dot{\Omega}$, we again prescribe the linear chirp $\Delta\omega(t)$, Eq.~\eqref{eq:ChirpWithB}, with a given constant ${\mathcal B}$. In this way, the inverse Bloch period
\begin{align}
	\frac{1}{t_\mathrm{B}} = \frac{\hbar}{8E_\mathrm{r}} \left({\mathcal B}-2l\dot{\Omega} \right)
	\label{eq:InvBlochTimeChirpRotation}
\end{align}
of the induced ABOs depends linearly on ${\mathcal B}$ with the offset $2l\dot{\Omega}$. This dependence is presented in Fig.~\ref{fig:Application2ExtAA}. We note that when the chirp constant ${\mathcal B}$ approaches the offset, that is when the angular velocity of external rotation is compensated by the chirp, and the resulting Bloch period becomes infinite.  

Thus, to measure the angular acceleration $\dot{\Omega}$, we perform two experiments with two sufficiently different values ${\mathcal B}_{1,2}$ of the chirp constant and measure the corresponding Bloch periods $t_\mathrm{B}^{(1,2)}$. According to Eq.~\eqref{eq:InvBlochTimeChirpRotation} and Fig.~\ref{fig:Application2ExtAA}, these two measurements completely determine the linear dependence of $t_{\rm B}^{-1}$ on $\mathcal B$ and the unknown angular acceleration is given by
\begin{align}
	\dot{\Omega} = \frac{1}{2l} \frac{{\mathcal B}_2 t_\mathrm{B}^{(2)}-{\mathcal B}_1 t_\mathrm{B}^{(1)}}{t_\mathrm{B}^{(2)}- t_\mathrm{B}^{(1)}}
	\label{eq:Application2OmegaDotDefinition}
\end{align}
in terms of prescribed ${\mathcal B}_{1,2}$ and measured $t_{\rm B}^{(1,2)}$.  

In conclusion, it is important to understand to which non-inertial force the proposed setup is in fact sensitive to. Rotation with a time-dependent angular velocity is described by the term $-\Omega_\mathrm{eff}(t)\hat{L}_z$ in Eq.~\eqref{eq:1DDynamicsMasterEquation} and corresponds to three non-inertial forces appearing in co-rotating frame, i.e. the Coriolis, centrifugal and Euler one. 
For atoms moving on the ring, both the Coriolis and the centrifugal forces act along the radial direction and are therefore compensated by the strong confinement. As a result, our scheme actually measures only the Euler force determined entirely by the angular acceleration $\dot{\Omega}$.

\section{Conclusion $\&$ Outlook} 
\label{sec:Conclusion}

We have predicted the new quantum phenomenon of the angular Bloch oscillations, which is analog of the ordinary Bloch oscillations in one dimension. For their observation we have proposed and analyzed in detail the experimental scheme based on state-of-the-art technologies used for ultra-cold quantum gases. In addition, we have presented valuable applications of ABOs to (i) transfer efficiently the angular momenta to atoms and (ii) measure precisely the angular acceleration of external rotation.

To realize the ABOs in a lab, we tightly confine atoms in a toroidal trap and add a ring lattice along the azimuth angle, made of two copropagating LG beams with the opposite angular indices $l$ and $-l$. We have demonstrated that in the presence of external rotation of small angular acceleration, or prescribed linear chirp between these LG beams, the measured angular momentum of trapped atoms displays the characteristic periodic behavior in time, i.e. ABOs. 

In the regime of a shallow ring lattice, ABOs can be utilized for the transfer of the well-defined {\it quantized} portion of angular momenta from the field to the trapped atoms. Moreover, to design a compact and reliable {\it quantum} sensor for the Euler non-inertial force, we have proposed an experimental scheme based on ABOs for measuring small angular acceleration of external rotation. It is important to mention that a similar setup, based on a pair of deep state-dependent counter-rotating ring lattices, has recently been proposed as Sagnac atom interferometer for measuring the constant angular velocity~\cite{Malinovsky2023}.

As outlook, our theoretical model of the proposed setup might be improved further by solving the 3D Schr\"odinger equation~\eqref{eq:Setup3DDynamics} numerically. This allows us to quantify the effect of coupling between radial and angular components, since this coupling puts limitations on the maximal angular momentum that can efficiently be transferred to the atoms. Moreover, it is also important to understand the effect of atom-atom interaction on the dynamics of ABOs and their applications.  

Furthermore, here we have analyzed the optical trapping using a bright lattice formed by two LG beams with the angular indices $l$ and $-l$. However, two LG beams with certain indices can create a dark lattice~\cite{FrankeArnold2007, Malinovsky2023}, resulting in reduction of photon scattering and hence longer coherence times. It might also be beneficial to consider other setups realizing a ring confinement with a additional periodic modulation along the angular variable, \textit{e.g.} by combining magnetic and optical traps~\cite{FrankeArnold2007,Amico2021,AtomtronicRMP2022}.

\begin{appendix}
    \section{Adiabatic time evolution}
    \label{sec:AppendixAdiabatic}
    
    Here, we investigate the adiabatic time evolution of our system, prepared initially in its ground state. 
    To do it, we first expand the solution 
    \begin{align}
    	\label{AppendixA:expension}
    	\Ket{\Theta(t)}=\sum_{n,m} c_{n,m}(t)\Ket{\Theta_{n,m}^t}
    \end{align}
    of the Schr\"odinger equation
    \begin{align}
    	\label{AppendixA:Schr equation}
    	\ii \hbar \frac{\partial}{\partial t}\Ket{\Theta(t)} = \hat{H}(t)\Ket{\Theta(t)}
    \end{align}
    with the time-dependent Hamiltonian
    \begin{align}
    	\label{AppendixA:Hamiltonian}
    	\hat{H}(t)= -\frac{\hbar^2}{2 I} \frac{\partial^2}{\partial\varphi^2} + V(t)\cos^2(l\varphi) +\Omega_{\rm eff}(t)\ii \hbar \frac{\partial}{\partial\varphi}
    \end{align}
    in terms of the Bloch states as instantaneous eigenstates
    \begin{align}
    	\label{AppendixA:functions u}
    	\Braket{\varphi | \Theta_{n,m}^t} = \ee^{\ii m \varphi} u_{n,m+\eta(t)}^{V(t)}(\varphi)
    \end{align}
    with the integer $m\in(-l,l]$ and $\eta(t)=-I\Omega_\mathrm{eff}(t)/\hbar$. 
    
    The functions $u_{n,q}^{v}(\varphi)$ are the solutions of the differential equation
    \begin{align}
    	{E}_n^v(q) u_{n,q}^v = 
    	\left[
    	\frac{1}{2 I} \left(-\ii \hbar\frac{\dd}{\dd\varphi} + \hbar q\right)^2+v \cos^2(l\varphi)
    	\right]
    	u_{n,q}^v
    	\label{AppendixA:Ustationarysolutions}
    \end{align}
    with $n=0,1,2\ldots$. 
    This equation should be solved for $-\pi/2l < \varphi \leq \pi/2l$ and with the periodic boundary conditions for $u_{n,q}^v(\varphi)$. 
    
    By inserting Eq.~\eqref{AppendixA:expension} into Eq.~\eqref{AppendixA:Schr equation} and using the eigenvalues ${E}^{\Theta}_{n,m} = {E}_n^{V}(m+\eta)-\hbar^2 \eta^2/(2I)$ of $\hat{H}(t)$, we arrive at the system of coupled ordinary differential equations
    \begin{align}
    	\begin{aligned}
    		\left[
    		\frac{\partial}{\partial t}
    		+
    		\frac{\ii}{\hbar} E_{n,m}^\Theta (t)
    		+
    		\Braket{\Theta_{n,m}^t | \frac{\partial}{\partial t} | \Theta_{n,m}^t}
    		\right]
    		c_{n,m}(t)
    		\\
    		=
    		\sumsum_{(n',m')\neq (n,m)} 
    		\frac{
    			\Braket{\Theta_{n,m}^t | \frac{\partial \hat{H}}{\partial t} | \Theta_{n',m'}^t}
    		}{
    			E_{n,m}^\Theta (t)-E_{n',m'}^\Theta (t)
    		}
    		c_{n',m'}(t)
    	\end{aligned}
    	\label{eq:ApendixAdiabaticCoefficients}
    \end{align}
    for the expansion coefficients $c_{n,m}(t)$.

    When the right-hand side of Eq.~\eqref{eq:ApendixAdiabaticCoefficients} is negligible, that is if
    \begin{align}
    	\label{AppendixA:Adiabaticity condition}
    	\hbar
    	\abs{\Braket{\Theta_{n,m}^t | \frac{\partial \hat{H}}{\partial t}| \Theta_{n',m'}^t}} \ll \Big|E_{n,m}^\Theta (t)-E_{n',m'}^\Theta (t))\Big|^2
    \end{align}
    with $(n',m')\neq (n,m)$, the system stays in the state, where it has been initially prepared. 
    This is the adiabatic limit. 
    
    As a result of adiabatic evolution, for our system prepared initially in its ground state, that is $c_{0,0}(0)=1$ and $c_{n,m}(0)=0$ for $n>0$ or $m>0$, we obtain   
    \begin{align}
    	\Theta(\varphi,t) 
    	= 
    	\ee^{\ii \phi(t)} \Theta_{0,0}^t(\varphi)
    	=
    	\ee^{\ii \phi(t)} u_{0, \eta(t)}^{V(t)}
    	,
    \end{align}
    where the imprinted time-dependent phase
    \begin{align}
    	\phi(t) = \phi_\mathrm{d}(t) + \phi_\mathrm{g}(t)
    \end{align}
    consists of the dynamical phase
    \begin{align}
    	\phi_\mathrm{d}(t) = 
    	-
    	\frac{1}{\hbar}
    	\int_{0}^t \! {E}_{0, 0 }^{\Theta} (s t')\, \dd t'
    	\label{AppendixA:DynamicalPhase}
    \end{align}
    and the geometrical phase
    \begin{align}
    	\phi_\mathrm{g}(t) = \ii \int_{0}^t \! 
    	\Bra{u_{0,\eta(t')}^{V(t')}} \frac{\partial}{\partial t'} \Ket{u_{0, \eta(t')}^{V(t')}}
    	\, \dd t',
    	\label{AppendixA:GeometricalPhase}
    \end{align}
    originating from the change of the instantaneous eigenstates $\Ket{u_{0, \eta(t)}^{V(t)}}$ with time \cite{Schleich}.

    \section{Adiabatic loading process}
    \label{sec:AppendixLoadingProcess}
    
    In this Appendix we study in detail the loading process, taking place at the second step of the scheme proposed for observation of ABOs and presented in Section~\ref{sec:Procedure} and in Fig.~\ref{fig:Procedure}. 
    Here, the external rotation as well as the chirp between the LG beams are absent, $\Omega(t)=0$ and $\Delta\omega(t)=0$, while the magnitude $V(t)$ of the ring lattice should be increased slowly, to ensure adiabaticity of the loading process. 
    For simplicity, we take a linear ramp function
    \begin{align}
    	V(t) =V
    	\begin{cases}
    		0, & t<-t_\mathrm{L}\\
    		(t+t_\mathrm{L})/t_\mathrm{L}, & -t_\mathrm{L}\leq t \leq 0\\
    		1, & 0<t,
    	\end{cases}
    \end{align}
    where $V$ is the final value of the magnitude $V(t)$ after the loading time $t_\mathrm{L}$.   
    
    According to Appendix~\ref{sec:AppendixAdiabatic}, the dynamics of the loading process is adiabatic if the condition given by Eq.~\eqref{AppendixA:Adiabaticity condition} is fulfilled. 
    Since $\partial\hat{H}(t)/\partial t=(\dd V(t)/\dd t)\cos^2(l\varphi)$, the matrix elements on the left-hand side of Eq.~\eqref{AppendixA:Adiabaticity condition} are not zero only for $m'=m$ and even values of $n'-n$. 
    Therefore, for our system prepared initially in the ground state of the toroidal trap, $\Theta (\varphi,-t_\mathrm{L})=u_{0,0}^0(\varphi)=1/\sqrt{2\pi}$, the most probable transition from $(n',m')=(0,0)$ is the one to the state $(n,m)=(2,0)$. 
    As a result, in the shallow lattice regime, $V\ll E_\mathrm{r}$, the adiabacity condition, Eq.~\eqref{AppendixA:Adiabaticity condition}, reads
    \begin{align}
    	\hbar 
    	\abs{\Braket{{\Theta^t_{2,0}}| \frac{\dd V}{\dd t}\cos^2(l\varphi)|\Theta^t_{0,0}}}\ll \abs{E_{2,0}^\Theta(t)-E_{0,0}^\Theta(t)}^2,
    \end{align}
    or
    \begin{align}
    	\left( \frac{ V}{E_\mathrm{r}}\right) \left( \frac{\hbar}{E_\mathrm{r} t_\mathrm{L}} \right) \ll 32\sqrt{2}
    	,
    	\label{eq:AdiabaticConditionLoading}
    \end{align}
    where we have approximated the states and the energy bands by the exact solutions for the case $V=0$. 
    
    To obtain the quantitative criterion for adiabacity, in other words, the inequality for the ratio $V/t_\mathrm{L}$, we have performed numerical simulations for the loading process. 
    For $-t_\mathrm{L}\leq t\leq 0$, we represent the state
    \begin{align}
    	\Ket{\Theta(t)} = \sum_{n,m}\mathcal{C}_{n,m}(t) \Ket{\Theta_{n,m}^{t=0}}
    \end{align}
    in terms of the stationary eigenstates $\Ket{\Theta_{n,m}^{t=0}}$ of the Hamiltonian $\hat{H}(0)$, Eq.~\eqref{AppendixA:Hamiltonian}, with $\Omega_\mathrm{eff}=0$. 
    By inserting this expansion into Eq.~\eqref{AppendixA:Schr equation}, we have derived the system of ordinary differential equations for the coefficients $\mathcal{C}_{n,m}(t)$ and solved it for the initial conditions $\mathcal{C}_{0,0}(-t_\mathrm{L})=1$ and $\mathcal{C}_{n,m}(-t_\mathrm{L})=0$ for $n>0$ or $m>0$.   
    
\begin{figure}
    	\centering
    	\includegraphics[width=\columnwidth]{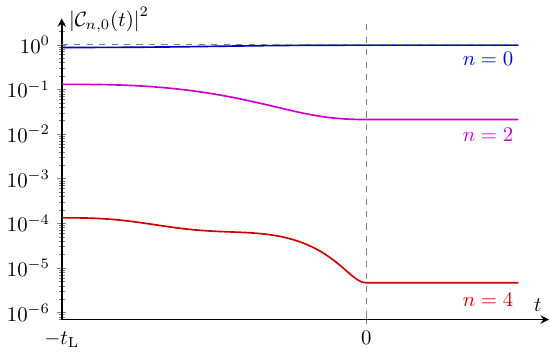}
    	\caption{
    		The population $\abs{\mathcal{C}_{n,0}(t)}^2$ of the Bloch states of the ring lattice with $n=0,2,4$ as functions of time $t$, for $V=5 E_\mathrm{r}$, $l=2$ and the loading time $t_\mathrm{L}=1/\omega_\mathrm{r}$.
    	} 
    	\label{fig:AppendLoadingCoefficients} 
    \end{figure}

    Fig.~\ref{fig:AppendLoadingCoefficients} displays the population $\abs{\mathcal{C}_{n,m}(t)}^2$ of the Bloch states of the ring lattice with $m=0$ and $n=0,2,4$ as functions of time $t$. 
    The results are shown for $V=5 E_\mathrm{r}$, $l=2$, and $t_\mathrm{L}=1/\omega_\mathrm{r}$. The population of the ground state $(0,0)$ increases during the loading process, while the populations of other states decrease.
    The loading process is finished at $t=0$, marked by the vertical dashed line in Fig.~\ref{fig:AppendLoadingCoefficients}. 
    The population of the state $(2,0)$ achieves a few percentages due to the non-adiabatic transitions occurring during the loading process. 
    
\begin{figure}[h]
    	\centering
    	\includegraphics[width=\columnwidth]{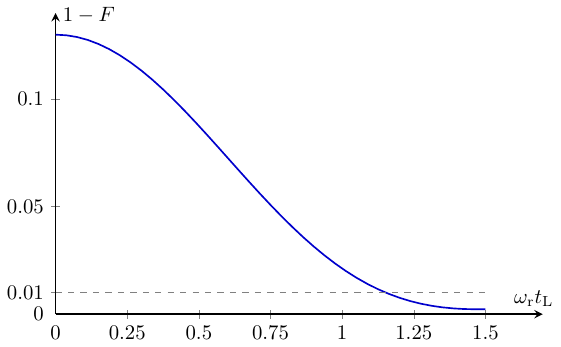}
    	\caption{
    		The fidelity $F$ of the loading processes as the function of the loading times $t_\mathrm{L}$, for $V=5 E_\mathrm{r}$ and $l=2$. The dashed line marks the fidelity $F=0.99$.
    	} 
    	\label{fig:AppendLoadingFidelity} 
    \end{figure}

    In addition, we have quantitatively analyzed the loading efficiency $F$, defined as the fidelity of the final (at $t=0$) and the target states $(0,0)$.
    For $V=5 E_\mathrm{r}$ and $l=2$, we present in Fig.~\ref{fig:AppendLoadingFidelity} the offset $1-F$ of the fidelity as the function of $t_\mathrm{L}$, where the dashed line identifies the level $F=0.99$, achieved for $t_\mathrm{L}\gtrsim 1.2/\omega_\mathrm{r}$.
    This result perfectly fits our adiabacity criterion, Eq.~\eqref{eq:AdiabaticConditionLoading}.
    
    \section{Angular acceleration and Landau-Zener transitions}
    \label{sec:AppendixLZ}

    Here we investigate the loss of atoms from the lowest energy band $E_0(q)$ to the higher energy band $E_1(q)$, depicted in Fig.~\ref{fig:EnergyDisp}, during the process of ABOs derived for a shallow lattice. 
    The dynamics of the atoms is governed by Eq.~\eqref{AppendixA:Schr equation} with the Hamiltonian $\hat{H}(t)$ given by Eq.~\eqref{AppendixA:Hamiltonian}, where $V$ is kept constant and $\Omega_\mathrm{eff}(t)$ is the prescribed function of time. 
    Moreover, our system is assumed to be initially prepared in its ground state $\Ket{\Theta_{0,0}^{t=0}}$. 
    
    When we try to use Eq.~\eqref{eq:ApendixAdiabaticCoefficients} to describe the dynamics of ABOs in the regime of shallow lattice, we find out that the right-hand side becomes quite large for certain $(n',m')$, leading to non-adiabatical dynamics. 
    Indeed, the matrix element
    $$
    \Braket{\Theta_{n,m}^t | \frac{\partial \hat{H}}{\partial t}| \Theta_{n',m'}^t}=-\frac{\dd \Omega_\mathrm{eff}}{\dd t} \Braket{{\Theta^t_{n,m}}| 
    	\hat{L}_z
    	|\Theta^t_{n',m'}}
    $$
    in Eq.~\eqref{eq:ApendixAdiabaticCoefficients} implies only transitions between different bands ($n'\neq n$) with $m'=m$. 
    Since our system is initially prepared in the $(0,0)$-state of the shallow lattice, the difference of the relevant time-dependent energies ${E}^{\Theta}_{0,0}(t)-{E}^{\Theta}_{n',0}(t)=E_0^{V}[\eta(t)]-E_{n'}^{V}[\eta(t)]$ in the denominator becomes very small as $n'=1$ and $\eta(t)=l$, that is when the quasi angular momentum $\eta(t)$ reaches the edge of the first Brillouin zone $\eta(t)=l$, as depicted in Fig.~\ref{fig:EnergyDisp}.   
    
    To examine quantitatively the non-adiabatic transitions in the shallow lattice, we reconsider Eq.~\eqref{AppendixA:Ustationarysolutions} in the limit $V\rightarrow 0$. At $V=0$, the solutions of this equation are given by $\ee^{- \ii 2l k \varphi}$, with $k=0,\pm 1,\pm 2,\ldots$, for any value of $q$ and correspond to the energies $\mathcal{E}_k(q) = (q-2lk)^2 E_\mathrm{r}/l^2$. 
    To define the bands, these solutions should be sorted according to their energies. Namely, for $q \in (0,l)$ the state with $k=0$ corresponds to the ground state band ($n=0$) and the state with $k=1$ to the first excited band ($n=1$), while in the range $q \in (l,2l)$ these roles are interchanged. Most importantly, at $q=l$ the energies of these two states are equal, that is the energy bands cross.
    
    For small $V>0$, this degeneracy is lifted. Following the standard methods of quantum mechanics, we can describe the dynamics of our system by writing the wave function
    \begin{align}
    	\label{eq:Linear_superposition}
    	\Theta (\varphi,t) = d_0 + d_1 \ee^{-\ii 2l\varphi}
    \end{align}
    in the form of the linear superposition of these two degenerate bands, or the non-perturbed torus states. By inserting Eq.~\eqref{eq:Linear_superposition} into Eq.~\eqref{AppendixA:Schr equation}, we arrive at the equations
    \begin{alignat}{1}
    	\ii \hbar \frac{\dd}{\dd t} {d}_{0} 
    	& =
    	\frac{V}{2}
    	d_{0} 
    	+
    	\frac{V}{4}
    	d_{1} 
    	\label{eq:DegPertTheoSystemofEq1} 
    	\\
    	\ii \hbar \frac{\dd}{\dd t} {d}_{1} 
    	& =
    	\left[
    	4E_\mathrm{r}
    	\left(
    	1-\frac{\eta(t)}{l}
    	\right)
    	+
    	\frac{V}{2}
    	\right]
    	d_{1} 
    	+
    	\frac{V}{4}
    	d_{0}
    	\label{eq:DegPertTheoSystemofEq2} 
    \end{alignat}
    for the coefficients $d_0$ and $d_1$, where $\eta(t)=-I\Omega_\mathrm{eff}(t)/\hbar$.
    
    For any time dependent $\eta(t)$ we can shift the common terms into the phase
    \begin{align}
    	\phi_0
    	= 
    	\frac{1}{\hbar}
    	\int_0^t 
    	\!
    	\dd t'
    	\left\{
    	2 E_\mathrm{r}
    	\left[
    	1-\frac{\eta(t')}{l}
    	\right]
    	+
    	\frac{V}{2}
    	\right\}
    	\label{eq:LZDegPertOffsetPhase}
    \end{align}
    and bring Eqs.~\eqref{eq:DegPertTheoSystemofEq1} and~\eqref{eq:DegPertTheoSystemofEq2} into symmetric form
    \begin{alignat}{3}
    	\ii \hbar \frac{\dd}{\dd t} D_0 
    	&= - &&2 E_\mathrm{r}
    	\left[
    	1-\frac{\eta(t)}{l}
    	\right] D_0
    	&&+
    	\frac{V}{4} D_1 \\
    	\ii \hbar \frac{\dd}{\dd t} D_1
    	&= +&&2 E_\mathrm{r}
    	\left[
    	1-\frac{\eta(t)}{l}
    	\right] D_1
    	&&+
    	\frac{V}{4} D_0,
    \end{alignat}
    where $D_{0}=\ee^{\ii\phi_0} d_{0}$ and $D_{1}=\ee^{\ii\phi_0} d_{1}$.
    
    For a linear behaviour $\eta(t) = st$, these two equations perfectly coincide with the well-known time evolution for the Landau-Zener process~\cite{Zener1932}. As a result, the transition probability, or the loss of atoms, 
    \begin{align}
    	T_\mathrm{LZ} = \exp{\left(-2\pi \gamma\right)}
    	\label{eq:LZTR}
    \end{align}
    from the lowest band to the first excited band during one passage through the Brillouin zone edge is determined by the Landau-Zener parameter~\cite{Zener1932}  
    \begin{align}
    	\gamma = \frac{1}{\hbar} \frac{(V/4)^2}{(4E_\mathrm{r} s/l)}.
    	\label{eq:LZgamma}
    \end{align}
    
    This approximation is valid when the effective interaction region, corresponding to a small energy difference, is small compared to the size of the  Brillouin zone. This is in fact the case for shallow lattices.
    
    In addition to the geometrical and the dynamical phase, the state that remains in the lowest band experiences the phase shift~\cite{Zener1932} 
    \begin{align}
    	\phi_\mathrm{LZ} = \frac{\pi}{4}+\arg \left[\Gamma\left( 1-\ii \gamma \right)\right]+\gamma \left[ \ln(\gamma)-1 \right]
    	\label{eq:LZPhiLZ}
    \end{align}
    due to the interaction with the higher band, where $\Gamma(\xi)$ denotes the Gamma function.
    
    For the given potential depth and small enough scanning speed $s$, the parameter $\gamma$ is large and the non-adiabatic losses are exponentially small, as shown by Eq.~\eqref{eq:LZTR}. Therefore, for $\gamma\gg 1$, we can use the identity $\Gamma\left( 1-\ii \gamma \right)= -\ii \gamma \Gamma\left( -\ii \gamma \right)$ and the Stirling's formula
    \begin{align}
    	\begin{aligned}
    		\ln\left[\Gamma\left( -\ii \gamma \right)\right]
    		=&
    		-\ii \gamma \ln \left(
    		-\ii \gamma
    		\right)
    		+
    		\ii \gamma
    		\\
    		&
    		+
    		\frac{1}{2} \ln \left(
    		\ii \frac{2\pi}{\gamma}
    		\right)
    		+
    		\mathcal{O}
    		\left(
    		\frac{1}{\gamma}
    		\right)
    	\end{aligned}
    \end{align}
    for the Gamma function, to obtain
    \begin{align}
    	\arg\left[\Gamma\left( 1-\ii \gamma \right)\right]
    	=
    	-\frac{\pi}{4}
    	+\gamma \left[1- \ln \left(
    	\gamma
    	\right)
    	\right]
    	+
    	\mathcal{O}
    	\left(
    	\frac{1}{\gamma}
    	\right)
    \end{align}
    and then
    \begin{align}
    	\phi_\mathrm{LZ} 
    	=
    	\mathcal{O}
    	\left(
    	\frac{1}{\gamma}
    	\right)
    \end{align}
    for the additional phase shift $\phi_\mathrm{LZ}$, Eq.~\eqref{eq:LZPhiLZ}.
    
    This limit reproduces perfectly the predictions of the adiabatic theorem: no transitions and no additional phase shift. Qualitatively, it takes place when $2\pi\gamma\ll 1$, or equivalently when $s\ll s_\mathrm{c}$ with the critical value
    \begin{align}
    	s_\mathrm{c} \equiv \frac{\pi l}{\hbar} \frac{V^2}{ 32 E_\mathrm{r}}.
    	\label{eq:DefSCrit}
    \end{align}
    
    To verify this estimation quantitatively, we employ the numerical simulations of Eqs.~\eqref{AppendixA:Schr equation}-\eqref{AppendixA:Hamiltonian} for $V=0.5\,E_\mathrm{r}$, $l=2$, and $s = 0.002\,\omega_\mathrm{r}$, corresponding to the critical value $s_\mathrm{c} \approx 0.049\,\omega_\mathrm{r}$ and accordingly $s\approx 0.04 s_\mathrm{c}$. In this case, 
    the mean value $\braket{\hat{L}_z}(t)$, given by Eq.~\eqref{eq:lzMeasurement} and presented in Fig.~\ref{fig:LargeAngularMomentumTransfer}, evidently displays  ABOs with $\braket{\hat{L}_z}(N_\mathrm{B}t_\mathrm{B})=-2l\hbar N_\mathrm{B}$, where $N_\mathrm{B}$ is the number of Bloch periods during the observation time $N_\mathrm{B}t_\mathrm{B}$.

\begin{figure}[t]
    	\centering
    	\includegraphics[width=\columnwidth]{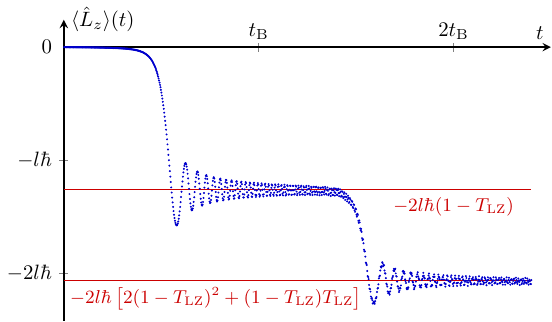}
    	\caption{
    		Mean angular momentum of the atoms in the case $V=0.5\,E_\mathrm{r}$,  $l=2$, and $s = s_\mathrm{c}=0.049\,\omega_\mathrm{r}$. The red lines are given by Eq.~\eqref{eq:NonAdiabaticTransitionsPartialTransitions}.
    	} 
    	\label{fig:AppendRotCrit} 
    \end{figure}

    However, for $s= s_\mathrm{c}$, the results are presented in Fig.~\ref{fig:AppendRotCrit} and clearly show the non-adiabatic transitions, as the steps in the function $\braket{\hat{L}_z}(t)$ do not reach the levels of the adiabatic dynamics, given by Eq.~\eqref{eq:ABOLzSteps}. 
    Instead, we have  
    \begin{align}
    	\begin{aligned}
    		\braket{\hat{L}_z}(N_\mathrm{B}t_\mathrm{B})
    		& = -2l\hbar
    		\Biggl[ N_\mathrm{B}(1-T_\mathrm{LZ})^{N_\mathrm{B}}\\ 
    		& +
    		T_\mathrm{LZ} \sum_{j=0}^{N_\mathrm{B}-1} j \left(1-T_\mathrm{LZ}\right)^{j} 
    		\Bigg]
    		\label{eq:NonAdiabaticTransitionsPartialTransitions}
    	\end{aligned}
    \end{align}
    and these values are displayed in Fig.~\ref{fig:AppendRotCrit} by the red horizontal lines. 
    
\begin{figure}[t]
    	\centering
    	\includegraphics[width=\columnwidth]{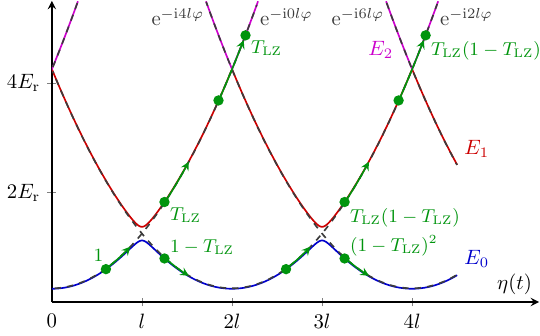}
    	\caption{
    		ABOs viewed with the energy bands in the limit of the shallow lattice. The green arrows indicate the Landau-Zener transitions between the bands and the green formulas give the populations of the corresponding non-perturbed torus states $\ee^{- \ii 2l k \varphi}$ with $k=0,1,2\ldots$, grey formulas. 
    	} 
    	\label{fig:AppendSeveralLZ} 
    \end{figure}

    To explain and derive these results, we use the band structure for a shallow lattice, Fig.~\ref{fig:AppendSeveralLZ}, and consider the dynamics of our system.
    For a shallow lattice we employ the first-order pertubation theory.
    Here the bands are well approximated by the parabolas $(q-2lk)^2 E_\mathrm{r}/l^2+V/2$ with $k=0,\pm 1,\pm 2,\ldots$, depicted by gray dashed lines and corresponding to the non-perturbed torus states $\ee^{- \ii 2l k \varphi}$. 
    However, at the edges of the Brillouin zones, the energy bands $E_0$ and $E_1$ have gaps of the size $V/2$, whereas gaps between all other neighbouring bands are negligible, e.g. the gap between $E_1$ and $E_2$ is $V^2/(32 E_\mathrm{r})$~\cite{Perrin2019}, as shown in Fig.~\ref{fig:AppendSeveralLZ}. 
    
    During the dynamics, the Landau-Zener transition occurs at the edge of the first Brillouin zone, $\eta=l$, and the fraction $T_\mathrm{LZ}$ of atoms is lost from the band $E_0$, blue line, to the higher band $E_1$, red line. Later, at $\eta=2l$, this fraction of atoms transits from the band $E_1$ to the band $E_2$, magenta line, and then to the high energy bands, as marked by the green arrow, because the gaps between the bands $E_n$ and $E_{n+1}$, with $n=1,2,\ldots$, are so small that the transition probability is almost one. However, this fraction $T_\mathrm{LZ}$ of atoms will always populate the state $\ee^{- \ii 0l \varphi}$. As a result, we obtain $\braket{\hat{L}_z}(t_\mathrm{B})=-2l\hbar (1-T_\mathrm{LZ})$ and this is the value marked by the red line in Fig.~\ref{fig:AppendRotCrit}. 
    
    At the edge of the second Brillouin zone, $\eta=3l$, there is next Landau-Zener transition between the bands $E_0$ and $E_1$ with the rate $T_\mathrm{LZ}$. Here only the fraction $T_\mathrm{LZ}(1-T_\mathrm{LZ})$ of atoms goes to the band $E_1$ and populates the state $\ee^{- \ii 2l \varphi}$, while the fraction $(1-T_\mathrm{LZ})^2$ of atoms remains in the band $E_0$ and populates the state $\ee^{- \ii 4l \varphi}$, as displayed in Fig.~\ref{fig:AppendSeveralLZ}. Hence, we have $\braket{\hat{L}_z}(2t_\mathrm{B})=-2l\hbar T_\mathrm{LZ}(1-T_\mathrm{LZ})-4l\hbar (1-T_\mathrm{LZ})^2$. This value is marked by the red line in Fig.~\ref{fig:AppendRotCrit}.  Analogously, by considering the next Landau-Zener transition at the edge of the third Brillouin zone, $\eta=5l$, and taking into account that the fractions $T_\mathrm{LZ}$, $T_\mathrm{LZ}(1-T_\mathrm{LZ})$, $T_\mathrm{LZ}(1-T_\mathrm{LZ})^2$, and $(1-T_\mathrm{LZ})^3$ of atoms populate the states $\ee^{- \ii 0l \varphi}$, $\ee^{- \ii 2l \varphi}$, $\ee^{- \ii 4l \varphi}$, and $\ee^{- \ii 6l \varphi}$, accordingly, we arrive at Eq.~\eqref{eq:NonAdiabaticTransitionsPartialTransitions} for $N_B=3$.   
    
    \section{Ground state energy band}
    \label{sec:AppendixApproximateGroundStateBand}
    
    In this Appendix, we derive the analytical formulas for the first two energy bands $E_0(q)$ and $E_1(q)$ for a shallow lattice. These help us to evaluate analytically the different parameters of the observable quantities utilized for the applications of ABOs.
    
    In Appendix~\ref{sec:AppendixLZ} we have already discussed the non-adiabatic dynamics of our system in the limit of a shallow lattice occurring around the edge of the first Brillouin zone $q=l$. By putting the constant quasi angular momentum $q$ instead of the time-dependent one $\eta(t)$ in Eqs.~\eqref{eq:DegPertTheoSystemofEq1} and \eqref{eq:DegPertTheoSystemofEq2} as well as using the relation Eq.~\eqref{eq:Energy_relation} between the eigenvalues of system \eqref{eq:DegPertTheoSystemofEq1}-\eqref{eq:DegPertTheoSystemofEq2} and the energies $E_n(q)$ of Eq.~\eqref{eq:Ustationarysolutions}, we arrive at
    \begin{align}
    	\begin{split}
    		E_{0}(q) & \approx 
    		\frac{V}{2}
    		+
    		E_\mathrm{r}
    		\left[
    		1+\left(1-\frac{q}{l}\right)^2
    		\right]
    		\\
    		& \phantom{\approx}
    		- 
    		\sqrt{
    			\frac{V^2}{16}
    			+
    			4 E^2_\mathrm{r} \left(1-\frac{q}{l}\right)^2 
    		}
    		\label{eq:ApproximationEnergyband0}
    	\end{split}
    \end{align}
    and
    \begin{align}
    	\begin{split}
    		E_{1}(q) & \approx 
    		\frac{V}{2}
    		+
    		E_\mathrm{r}
    		\left[
    		1+\left(1-\frac{q}{l}\right)^2
    		\right]
    		\\
    		& \phantom{\approx}
    		+ 
    		\sqrt{
    			\frac{V^2}{16}
    			+
    			4 E^2_\mathrm{r} \left(1-\frac{q}{l}\right)^2 
    		},
    		\label{eq:ApproximationEnergyband1}
    	\end{split}    
    \end{align}
    valid for $|q-l|\ll l$.  
    
\begin{figure}
    	\centering
    	\includegraphics[width=\columnwidth]{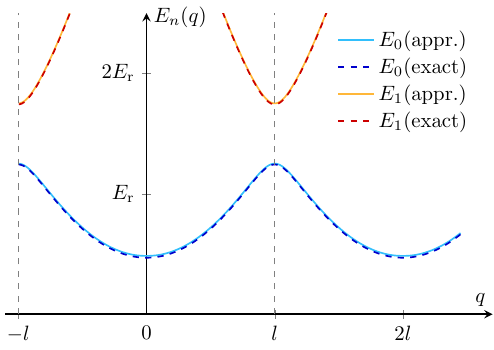}
    	\caption{
    		Comparison of the energy bands $E_0(q)$ and $E_1(q)$ obtained with solving Eq.~\eqref{eq:Ustationarysolutions} numerically for $V=E_\mathrm{r}$, dashed lines, to the ones given by  Eqs.~\eqref{eq:ApproximationEnergyband0} and \eqref{eq:ApproximationEnergyband1}, solid lines.
    	} 
    	\label{fig:AppendEnergyApprox} 
    \end{figure}

    In Fig.~\ref{fig:AppendEnergyApprox} we compare these results, presented by solid lines, to the exact ones, dashed lines, obtained with the numerical solution of Eq.~\eqref{eq:Ustationarysolutions} for $V=E_r$.  
    The bands are in fact well reproduced over the whole range of quasi angular momentum, $q\in(0,2l)$, and it justifies the validity of considering only the two unperturbed states of the system in the case of a shallow lattice. Thus, the whole band structure can be obtained by repeating $E_0(q)$ and $E_1(q)$, Eqs.~\eqref{eq:ApproximationEnergyband0} and \eqref{eq:ApproximationEnergyband1}, to the next Brillouin zones. 
    
    Now we apply Eq.~\eqref{eq:ApproximationEnergyband0} for deriving analytical formulas for the observable $\braket{\hat{L}_z}(t)$, Eq.~\eqref{eq:lzMeasurement}, in the regime of shallow lattice. By calculating the first derivative $\dd E_0/\dd q$ and putting $q=\eta(t)$, we obtain
    \begin{align}
    	\label{Appendix:Lz}
    	\braket{\hat{L}_z}(t) 
    	= -\hbar l\left[
    	1+
    	\frac{\eta(t)-l}{\sqrt{\left[\eta(t)-l\right]^2+\left[Vl/(8E_\mathrm{r})\right]^2}}
    	\right]
    \end{align}
    and
    \begin{align}
    	\label{Appendix:dLz}
    	\frac{\dd\braket{\hat{L}_z}}{\dd t}= -\hbar l\frac{\dd\eta}{\dd t}
    	\frac{\left[Vl/(8E_\mathrm{r})\right]^2}{\left\{\left[\eta(t)-l\right]^2+\left[Vl/(8E_\mathrm{r})\right]^2\right\}^{3/2}}.
    \end{align}
    
    For the linear function $\eta(t)=st$, Eq.~\eqref{eq:DefEta}, with the scanning parameter $s=I\mathcal{B}/(2\hbar l)$, corresponding to the case of no external rotation, $\Omega(t)=0$, and a linear frequency chirp $\Delta\omega(t)=-\mathcal{B}t$, Section~\ref{sec:Parameters}, the function $\dd\braket{\hat{L}_z}/\dd t$ has peaks at half-integer number of the Bloch period, $t_n=(n+1/2)t_\mathrm{B}$ with $n=0,1,2\ldots$, with heights
    \begin{align}
    	\left.
    	\frac{\dd \braket{\hat{L}_z}}{\dd t} 
    	\right|_{t=t_n}=
    	- 2l \frac{\hbar^2 \mathcal{B}}{V}
    	\equiv
    	- \mathcal{A},
    \end{align}
    as shown in Fig.~\ref{fig:ABODeriv}.
    
    Moreover, these peaks are sharp, since the their heights scale as $V^{-1}$, whereas their relative widths
    \begin{align}
    	\frac{\Delta t_\mathrm{FWHM}}{t_\mathrm{B}} = \frac{V}{8 E_\mathrm{r}}\sqrt{4^{1/3}-1}
    \end{align}
    scale as $V$ and are small for a shallow lattice, $V\ll E_r$. Here $\Delta t_\mathrm{FWHM}$ denotes the full width at half maximum of the peaks. 
    
\end{appendix}

\section*{Acknowledgment}
We are grateful to Eric Glasbrenner and Vladimir Malinovsky for the fruitful discussions on the Bloch oscillations and rotation sensing.

\bibliography{ABO_references}

\newpage

\end{document}